\documentclass[12pt]{article}
\usepackage{graphicx}
\usepackage{amsmath,amssymb,amscd}

\newcommand\pubnumber{DPF2015-165}
\newcommand\pubdate{October 28, 2015}

\def\Title#1{\begin{center} {\Large #1 } \end{center}}
\def\Author#1{\begin{center}{ \sc #1} \end{center}}
\def\Address#1{\begin{center}{ \it #1} \end{center}}

\newcommand\pubblock{\rightline{\begin{tabular}{l} \pubnumber\\
         \pubdate  \end{tabular}}}
\newenvironment{Abstract}{\begin{quotation}  }{\end{quotation}}
\newenvironment{Presented}{\begin{quotation} \begin{center} 
             PRESENTED AT\end{center}\bigskip 
      \begin{center}\begin{large}}{\end{large}\end{center} \end{quotation}}
\def\Acknowledgments{\bigskip  \bigskip \begin{center} \begin{large}
             \bf ACKNOWLEDGMENTS \end{large}\end{center}}

\def\beq{\begin{equation}}
\def\eeq#1{\label{#1}\end{equation}}
\def\eeqn{\end{equation}}

\def\beqa{\begin{eqnarray}}
\def\eeqa#1{\label{#1}\end{eqnarray}}
\def\eeqan{\end{eqnarray}}

\let\bar=\overbar

\def\Dslash{\not{\hbox{\kern-4pt $D$}}}
\def\dslash{\not{\hbox{\kern-2pt $\del$}}}

\def\msb{{\bar{\ssstyle M \kern -1pt S}}}

\begin{document}
\begin{titlepage}
\pubblock

\vfill
\Title{How scalar-field dark matter may conspire to facilitate baryogenesis at the electroweak scale}
\vfill
\Author{T. Rindler-Daller$^{1}$, B. Li$^{2}$, P.R. Shapiro$^{2}$, M. Lewicki$^{3}$, J.D. Wells$^{1}$}
\Address{$^{1}$Department of Physics, University of Michigan, Ann Arbor, MI 48109, USA\\
$^{2}$Department of Astronomy, University of Texas, Austin, TX 78712, USA\\
$^{3}$Institute of Theoretical Physics, University of Warsaw, Warsaw 02-093, Poland}
\vfill
\begin{Abstract}
 The cosmic evolution of a dark matter model which behaves relativistically in the early Universe is explored.
 Dark matter is described as a complex scalar field, whose earliest evolution is characterized by a stiff
 equation of state ($p \simeq \rho$). In this phase, it is the dominant component in the Universe.
 We present constraints from Big Bang nucleosynthesis and primordial gravity waves from inflation.
 Also, we study how the associated enhanced expansion rate due to the stiff phase might facilitate 
 a first-order electroweak symmetry breaking phase transition, in light of the recently measured value of the Higgs boson mass. 
\end{Abstract}
\vfill
\begin{Presented}
DPF 2015\\
The Meeting of the American Physical Society\\
Division of Particles and Fields\\
Ann Arbor, Michigan, August 4--8, 2015\\
\end{Presented}
\vfill
\end{titlepage}

\section{Dark matter as a complex scalar field: SFDM}

{\small
The standard model of particle physics is insufficient to explain two a-priori unrelated problems in modern cosmology, 
the nature of the cosmological dark matter (DM) and baryon asymmetry. The latter has long been sought to
be explained by baryogenesis at the electroweak scale. This has been proven to be very difficult
for a Higgs boson mass as high as 125 GeV, for which the electroweak phase transition is not sufficiently strong in a
standard radiation-dominated Universe. 
In this paper, we present a scenario which may solve these two problems at once. 
The crucial role is thereby played by the adopted nature of the DM. 
In our model, DM is not made of WIMPs, but of
ultralight bosons with masses of the order $10^{-21}$ eV.
Examples of ultralight bosons as DM candidates include those
predicted by multidimensional extensions to the standard model, be
they in the form of a hierarchy of axion-like particles in string
theory or of ``excitons'' in extradimensional cosmologies, akin to
Kaluza-Klein modes.
The respective origins and evolutionary histories of these
diverse models of ultralight bosonic DM may differ considerably, however.  
We are interested in scenarios in which these DM bosons find themselves rapidly occupying their ground state in 
the wake of inflaton decay and reheating, 
while the thermal contribution of DM bosons and antibosons is rapidly annihilated away.
(An example of such a microphysical implementation can be found in \cite{Mangano}.) 
As a result, the DM can thereafter be described as a classical field with a conserved U(1)-charge,
which is effectively the conserved DM abundance.
We have studied the cosmic evolution of this complex scalar field dark matter (SFDM) 
in the past and will outline some of our previous results in Section 2 of this paper. 
In particular, SFDM behaves like a relativistic fluid in the early Universe, where its contribution to $N_{\rm{eff}}$ - 
the number of relativistic degrees of freedom - has to be properly contrained by the time of matter-radiation-equality, 
Big Bang nucleosynthesis (BBN) and 
the primordial gravitational wave background from inflation. The first two constraints will be discussed in Section 2,
while new results of the effect of the latter will be discussed separately in Section 3. 
Thus, the dynamical evolution of SFDM giving rise to an increased expansion rate in the early Universe leads
to a host of novel implications for its evolution, including phase transitions of various kinds. Work is
in progress to study the implication of this model on the feasibility of a first-order electroweak phase
transition in the framework of the standard model, as well as in the standard model with a cutoff scale in which
the Higgs potential is augmented by a $\phi^6$-operator. Some results will be presented in Section 4. 
In the late Universe, SFDM must act like collisionless cold dark matter (CDM) in order to reproduce the successes of the latter.
It turns out that SFDM not only can reproduce CDM on large scales, but that it potentially resolves the problems of CDM on (small)
galactic scales:  depending on the parameters of the ultralight bosons of SFDM, the associated Jeans scale of gravitational collapse
can be of the order of a kpc, which is much larger than that of WIMPs. That Jeans
scale is either related to the particle de-Broglie wavelength, or is related to the pressure due to repulsive particle self-interactions, 
which can prevent collapse below kpc scales, as well. In either case, the formation of structure (hence galaxies), as well as the 
accumulation of DM in the centers of galaxies is prevented below those scales. Therefore, SFDM does not predict cuspy cores or an overabundance of
satellite galaxies, in contrast to CDM, whose opposite predictions have been continuously challenged by galaxy observations. 
There is thus an important astrophysical motivation to consider SFDM models with
ultralight mass, and implications have been studied by some of us and many others, see \cite{RS, RS2}
and the numerous references therein.}

\section{The cosmic evolution of $\Lambda$SFDM}

{\small Like for any scalar field, the evolution of SFDM is determined by the form of the potential in its Langrangian.
Let $\psi$ be the field
describing the condensate of DM bosons, we adopt the following Lagrangian (in units of energy density)
\begin{equation} \label{Lag}
\mathcal{L} = \frac{\hbar^2}{2m}g^{\mu \nu}\partial_{\mu}\psi^*\partial_{\nu}\psi - \frac{1}{2}mc^2|\psi|^2 - 
\frac{\lambda}{2}|\psi|^4, 
\end{equation}
with signature $(+,-,-,-)$. 
$m$ is the DM boson mass and the energy-independent boson coupling strength is chosen to be repulsive or zero, $\lambda \geq 0$.
The SFDM parameters of interest are tiny. A de-Broglie wave length of order kpc results from particles with mass
$m \simeq 10^{-22}$ eV$/c^2$. The small-scale cutoff problem of structure formation can only be resolved for bigger masses,
\textit{if} at least a tiny coupling is included. Fiducial 
dimensional couplings of order $\lambda \approx 10 ^{-62}$ eV cm$^3$ correspond to dimensionless couplings of
order $\lambda m^2 c/\hbar^3 \approx 10^{-86}$. It turns out that couplings even this small are enough to resolve the small-scale problems, but they
do render these models qualitatively different from those with $\lambda = 0$ in the early Universe.
The (quadratic) mass term in equ.(\ref{Lag}) guarantees that SFDM behaves like CDM in the late Universe. 
More precisely, this term must dominate after
the time of matter-radiation equality at a scale factor of $a \simeq 3\times 10^{-4}$, in order to reproduce an epoch of ``CDM-like''-domination.
Earlier in the evolution when the quartic term in (\ref{Lag}) dominates, 
the equation-of-state (EOS) of SFDM is that of radiation, namely $p_{\rm{SFDM}} \simeq \rho_{\rm{SFDM}}/3$. 
Going further backwards in time, it is the kinetic term in equ.(\ref{Lag}) that will dominate, 
and the EOS of SFDM will approach that of
maximal ``stiff'' matter, $p_{\rm{SFDM}} \simeq \rho_{\rm{SFDM}}$. A scalar-field with a stiff EOS has sometimes been
called ``fast-roll'' or ``kination'' in the literature. It is important to note that this earliest phase of SFDM
appears in all models, with or without $\lambda$. 

We studied this evolution in detail
in Ref.\cite{LRS} by solving numerically the equations-of-motion of SFDM in an expanding Universe, i.e. the Klein-Gordon equation
for $\psi$ in an expanding Friedmann-Robertson-Walker background. We termed this model $\Lambda$SFDM, since all the cosmic components
of the $\Lambda$CDM model are adopted, except for CDM which is replaced by SFDM. The present cosmic DM abundance is thus assumed to be entirely
given by the corresponding energy density of SFDM.
We found that, in the very early Universe, SFDM dominates the cosmic energy budget in its ``stiff'' phase. 
As the Universe expands further, SFDM transitions to
its radiation-like phase. However, it is the radiation component, which then dominates the cosmic energy budget, i.e.
a radiation-dominated epoch follows the epoch of SFDM-domination. Finally,
SFDM will transition to its CDM-like phase, where, again, it dominates and gives rise to matter-domination. That epoch is only
followed by $\Lambda$-domination, assuming a cosmological constant as in $\Lambda$CDM (see left-hand plot of Figure \ref{fig1}).
Now, in the stiff and radiation-like phases of SFDM, it contributes to $N_{\rm{eff}}$ - the number of relativistic degrees of freedom. 
$N_{\rm{eff}}$ can be probed by several cosmological observables, notably the time of matter-radiation equality
$a_{\rm{eq}}$ and BBN. At the time of $a_{\rm{eq}}$, SFDM must have morphed into a non-relativistic component. On the other hand,
BBN must proceed in a radiation-dominated background, i.e. the stiff phase of SFDM must end before BBN. We define the epoch
of BBN by two events: the time of neutron-proton freeze-out $a_{\rm{n/p}}$ followed by the time of nuclei production $a_{\rm{nuc}}$.
We require that SFDM must have fully morphed to its radiation-like phase by the time of $a_{\rm{nuc}}$, in such a way that the
evolution of $N_{\rm{eff}}$ of SFDM is in accordance with the $1\sigma$ uncertainty of current BBN constraints.
In \cite{LRS}, we used the 2013 Planck data release \cite{Planck13}, as well as $N_{\rm{eff}}=3.71^{+0.47}_{-0.45},$ from
BBN contraints of \cite{Steigman},
in order to constrain the allowed parameter space of SFDM models. It turned out that the resulting constraints were much tighter
than found in the literature previously. The bounds from $a_{\rm{eq}}$ and BBN obtained were $m\ge2.4\times10^{-21}$eV$/c^2$ and 
$9.5\times10^{-19}\rm{eV^{-1}cm^3}\le\lambda/(mc^2)^2\le4\times10^{-17}\rm{eV^{-1}cm^3}$. We stress that, by assuming that the 
$N_{\rm{eff}}$ due to SFDM must lie within the above $1\sigma$ bound of BBN, the latter inequality
excluded models with $\lambda=0$. (N.B. however, if future BBN abundances bring $N_{\rm{eff}}$ down, sufficiently close to
the standard value of $N_{\rm{eff}} = 3.046$, this may allow $\lambda = 0$). }

\section{$\Lambda$SFDM and primordial gravity waves\\ from inflation}

\begin{figure*}[h]
     \centering
     \includegraphics[width=7.5cm]{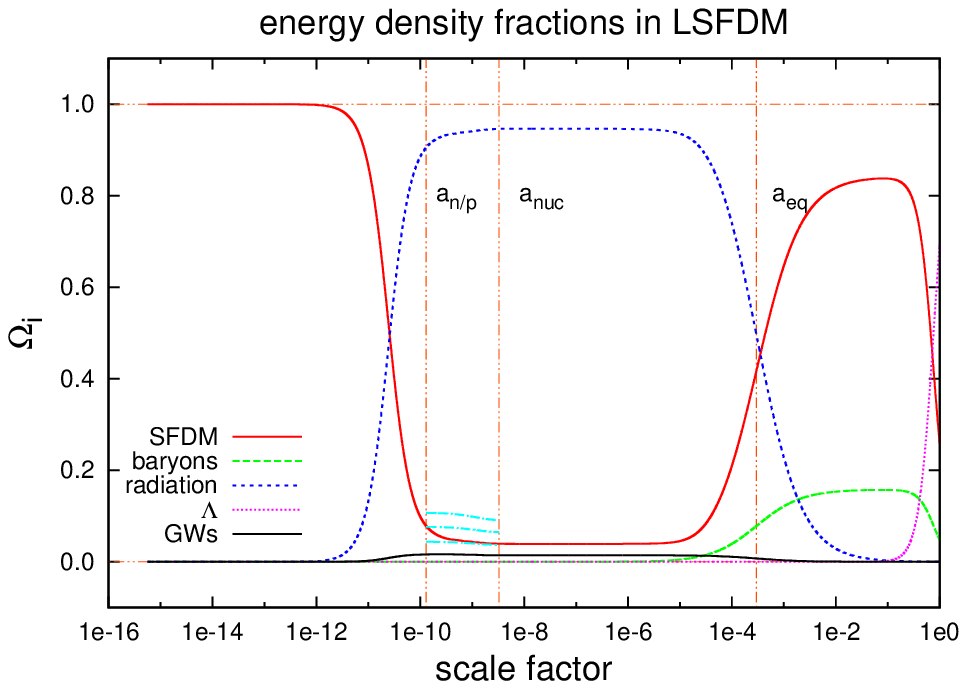}
      \includegraphics[width=7.5cm]{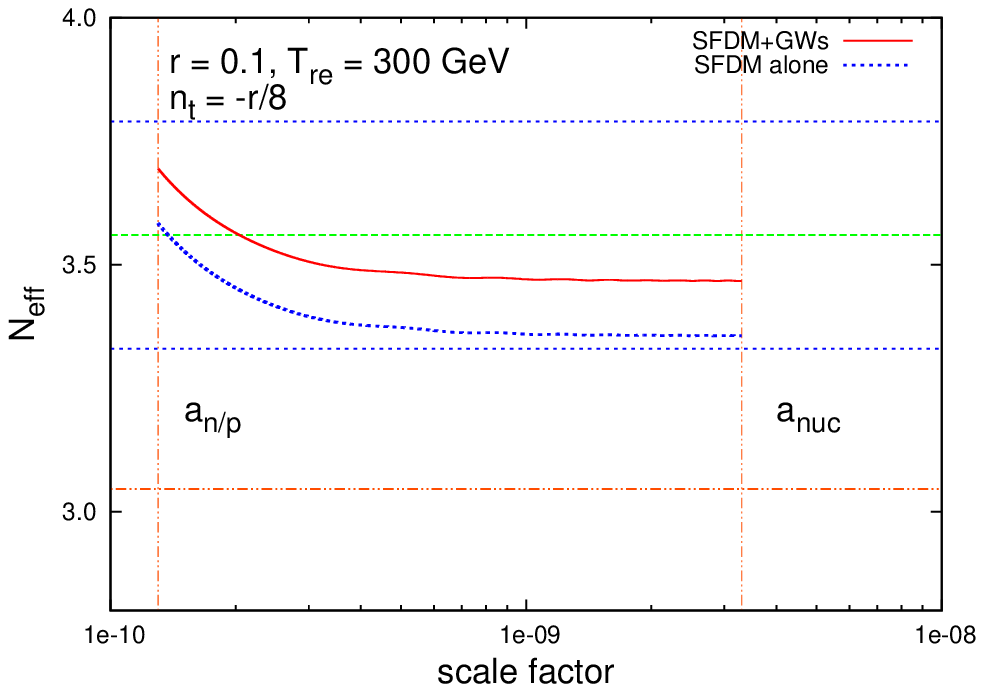}
   \caption{{\small Fiducial SFDM model: cosmic evolution of energy density fractions (left); evolution of $N_{\rm{eff}}$ during BBN
   for SFDM with and without gravity waves (right).}}
 \label{fig1}
  \end{figure*}

\begin{figure*}
\centering
\includegraphics[width=7.5cm]{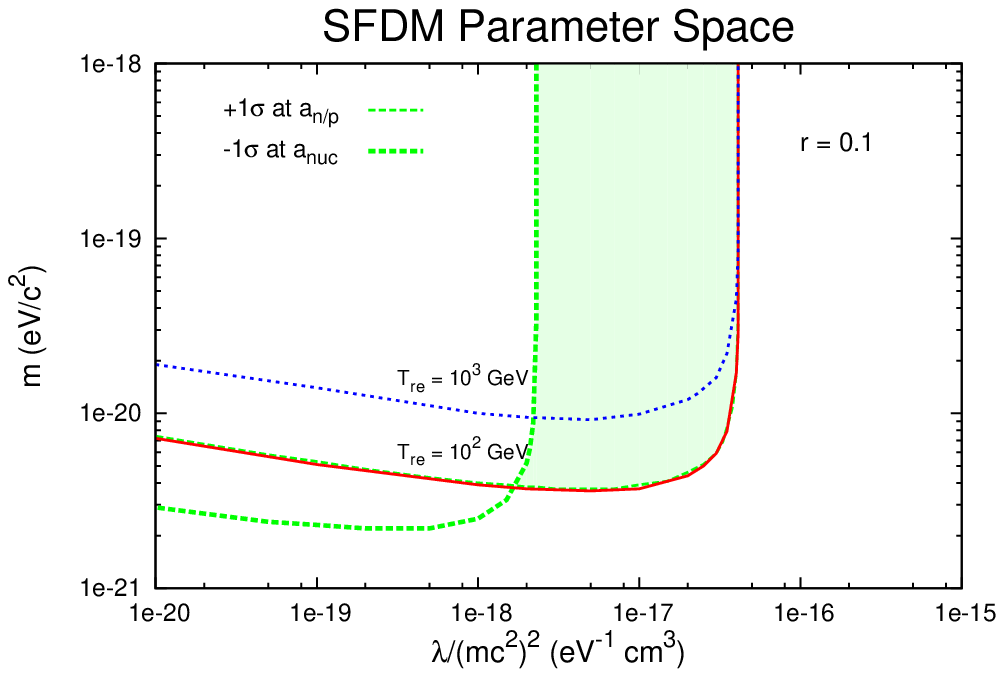}
\includegraphics[width=7.5cm]{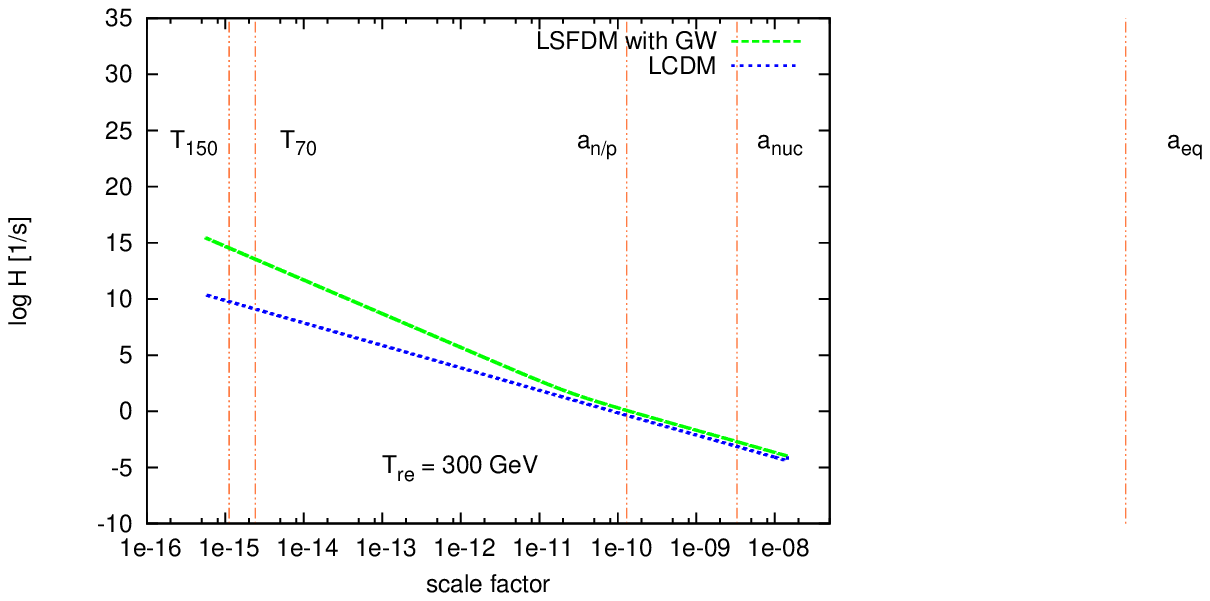}
\caption{{\small SFDM parameter space (left); Hubble expansion rate during EWSB and BBN for our fiducial $\Lambda$SFDM model and
$\Lambda$CDM, respectively (right).}}
\label{fig2}
\end{figure*}

{\small Lately, probing the amount of primordial gravity waves from inflation - 
which is an \textit{additional} source of $N_{\rm{eff}}$ to any other relativistic background - 
has become possible by advanced cosmic microwave background observations. In our latest work, 
we use the 2015 data
release of Planck \cite{Planck15}, along with an updated BBN value of $N_{\rm{eff}} = 3.56 \pm 0.23$ from \cite{NS} 
in order to contrain SFDM even further. A detailed analysis will be published elsewhere.
We assume that the primordial power spectrum of tensor fluctuations is generated during inflation and parameterized by a power law,
in terms of the (scalar) amplitude, the tensor-to-scalar ratio $r$, and the tensor spectral index $n_t$. 
This provides us with our initial cosmic amount
of gravitational waves. Then, we calculate the evolution of a $\Lambda$SFDM universe, including $\Omega_{\rm{GW}}$.

It has been realized since the early work of \cite{Grish}, that any amount of primordial $\Omega_{\rm{GW}}$ 
(which behaves as a radiation-like component with $p=\rho/3$) 
will be enhanced for a background EOS which is stiffer than radiation, i.e. for $p/\rho = w > 1/3$. In particular, the enhancement
is maximal for
a stiff EOS. Therefore, the constraint on SFDM from $N_{\rm{eff}}$ has two sources: the direct contribution from SFDM, 
and a new one from the \textit{enhanced} $\Omega_{\rm{GW}}$, which results 
from the presence of the stiff phase of SFDM.
Once SFDM transitions to radiation-like, the boost of $\Omega_{\rm{GW}}$ due to the stiff phase ends, and   
$\Omega_{\rm{GW}}$ thereafter only grows as it would in a standard radiation-dominated universe. 
More precisely, our cosmic evolution proceeds as follows: we assume that a de-Sitter-like inflationary phase
is followed by an epoch of reheating with $w = 0$. The DM bosons are born at the end of reheating and move into
their ground state, after which SFDM dominates the Universe in its stiff phase with $w=1$.  We adopt an instant transition, hence 
the reheating temperature $T_{\rm{re}}$ at the end of reheating also corresponds to the point after which there is the 
``stiff'' epoch. It turns out that the primordial $\Omega_{\rm{GW}}$ gets its biggest boost just after $T_{\rm{re}}$, when
modes re-enter the horizon at the onset of the stiff phase. Unsurprisingly, the overall enhancement is larger for higher $T_{\rm{re}}$, i.e. shorter
epochs of reheating, since the stiff phase is then correspondingly longer. This puts constraints on the SFDM model parameters
$m$ and $\lambda/m^2$. In the left-hand plot of Figure \ref{fig1}, we show the cosmological evolution for a 
\textit{fiducial SFDM model} with 
$m = 5\times 10^{-21}$ eV$/c^2$, $\lambda/(mc^2)^2 = 3\times 10^{-18}~ \rm{eV^{-1}cm^3}$, which fulfills 
the contraints from $a_{\rm{eq}}$, BBN, and $\Omega_{\rm{GW}}$, choosing a reheating temperature of $300$ GeV.
The different cosmic components evolve as a function of scale factor, as labelled in the legend. 
The right-hand plot shows the evolution of $N_{\rm{eff}}$ as a function of scale factor during BBN, 
for SFDM with or without $\Omega_{\rm{GW}}$ included. 
Clearly, not all SFDM models will be in accordance with the (current) bound on $\Omega_{\rm{GW}}$ plus $N_{\rm{eff}}$ from BBN. 
This can be also seen in the left-hand plot of Figure \ref{fig2}, which shows the allowed parameter space (filled area) for
two choices of reheating temperatures: the higher $T_{\rm{re}}$, 
the higher an SFDM particle mass is required
for the model to be in accordance with a given value of $r$ (we took $r = 0.1$). Following the consistency relation of standard slow-roll inflation,
this also implies a given value of the tensor spectral index $n_t$, according to $n_t = -r/8$. 
Roughly, a factor of 10 increase in $T_{\rm{re}}$ shifts the lower bound on mass upwards by a factor of 3-4. We can also see that the impact 
of different reheating temperatures on the allowed range of $\lambda/m^2$ is small. 
Indeed, the contraint on $\lambda/m^2$ comes mostly from the allowed value of $N_{\rm{eff}}$ during BBN. Since we chose a smaller
BBN $N_{\rm{eff}}$ than in our previous work, the allowed range of $\lambda/m^2$ is smaller, as a result.  }

\section{Implication of a $\Lambda$SFDM universe on EWBG}

\begin{figure*}[h]
\centering
\includegraphics[width=5.5cm]{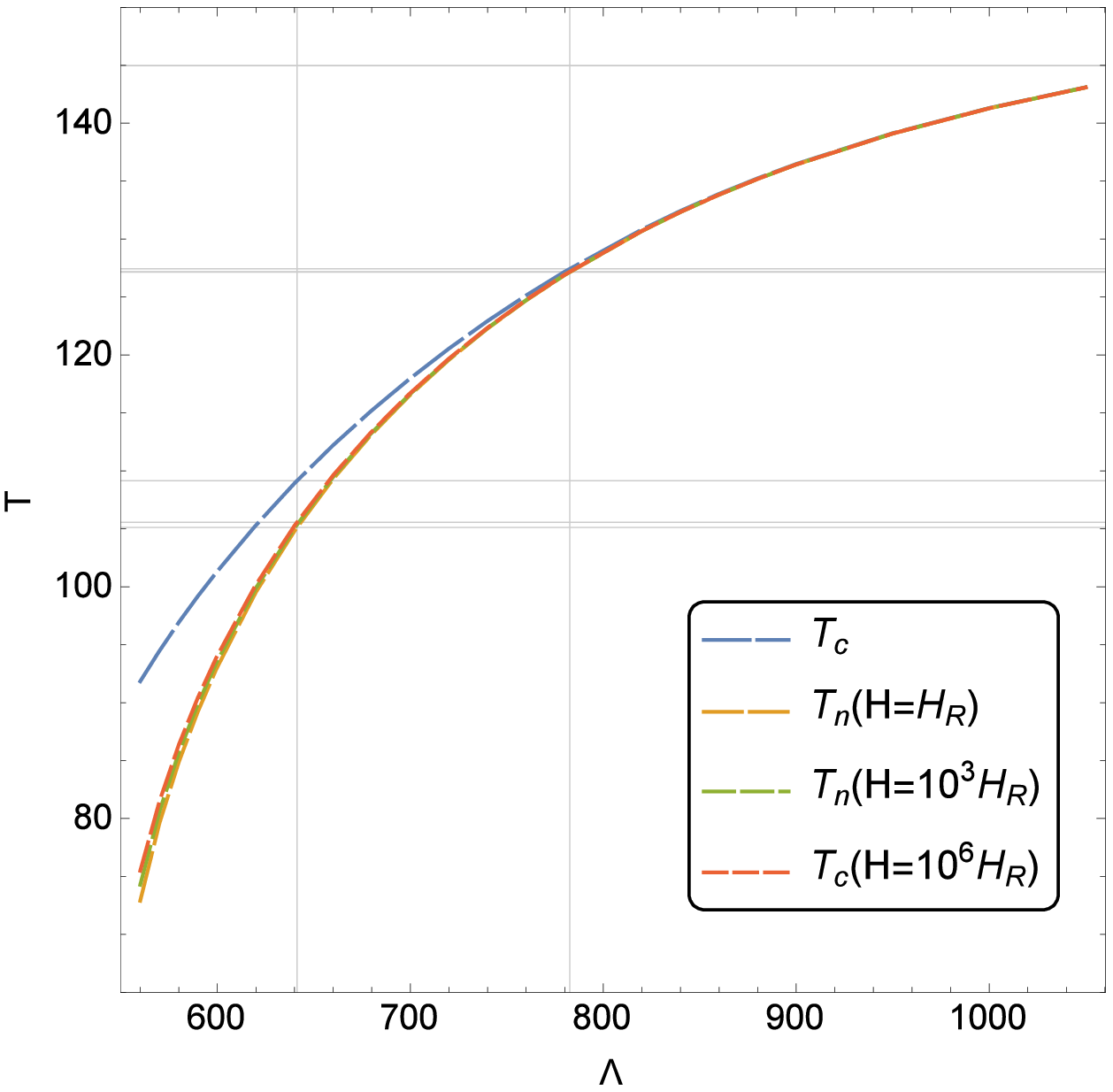}
\includegraphics[width=5.5cm]{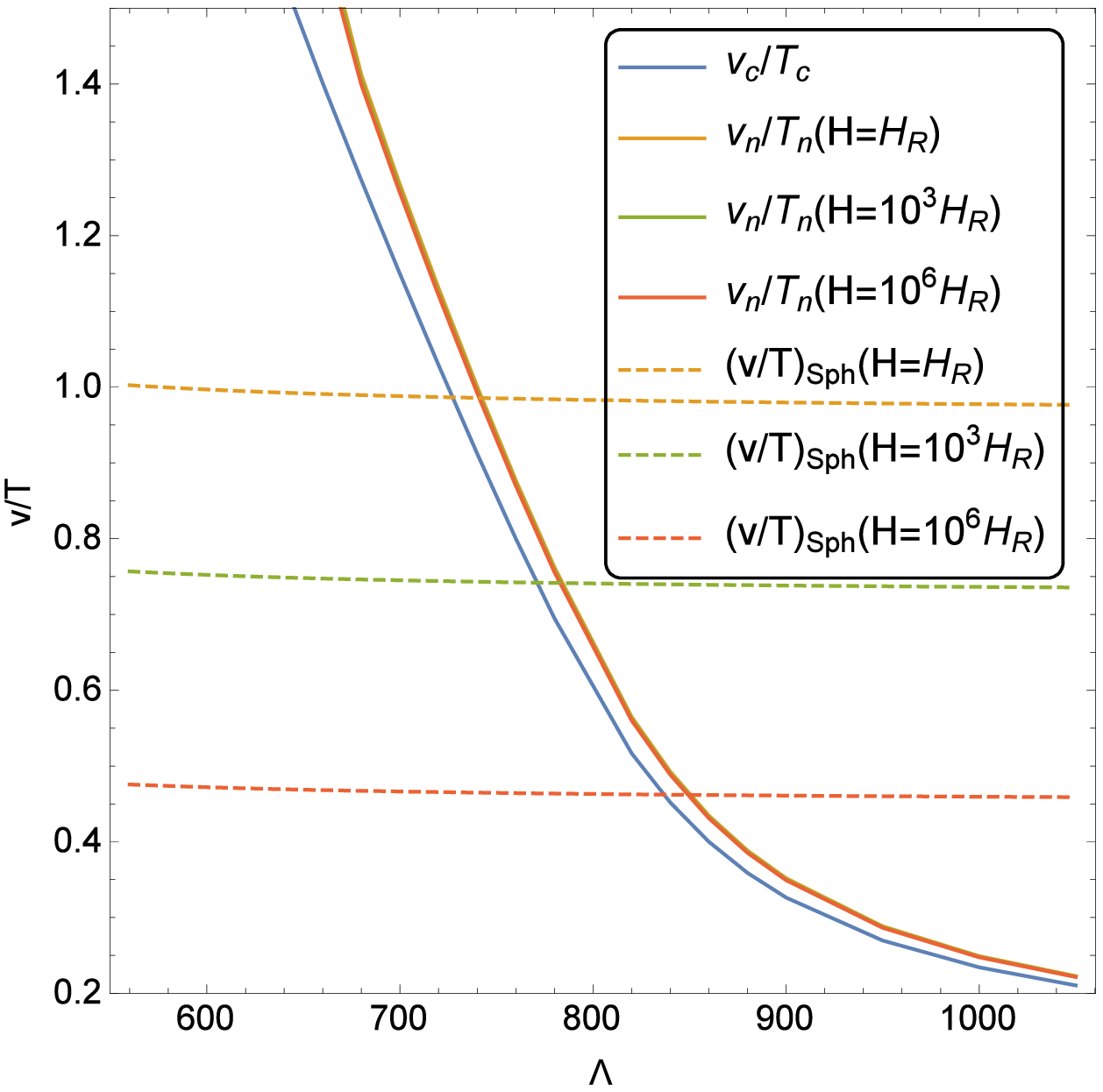}
\caption{{\small Critical and nucleation temperatures (left) and the ratio of Higgs vev to those temperatures (right) 
versus cutoff scale;
also indicated on the right are the bounds for sphaleron freeze-out for different expansion rates ($H_R$ denotes the expansion rate in 
$\Lambda$CDM).}}
\label{fig3}
\end{figure*}

{\small
A Universe with complex SFDM has a higher expansion rate during its ``stiff'' phase of SFDM-domination ($H \propto a^{-3}$), 
than it would have in a purely radiation-dominated phase ($H \propto a^{-2}$). This has implications for various phase transitions in the early Universe, 
which are usually assumed to happen during radiation-domination. The effect of a modified expansion history on various processes
has been studied in previous literature. One particular model of interest to us is that of \cite{Joyce} and \cite{JP}, where the 
presence of a ``fast rolling'' scalar field (``kination'') had been found to facilitate the creation of a baryon asymmetry,
even if the electroweak symmetry breaking (EWSB) phase transition (PT) is not or only very weakly of 1st-order.
The higher expansion rate results effectively in a weaker bound on the corresponding 
sphaleron freeze-out (or ``wash-out'') condition, which is equivalent to alleviating the strenght of the 1st-order PT.
The scalar field adopted there is either the inflaton itself, which transitions into the required kination phase, or is thought
to be some other relic (real) field with appropriate potential. In each case, the field energy density always decays as 
$\rho \propto a^{-6}$, and so its influence ceases
rapidly. Hence, this phase must be tailored to last during the EWSB, which, in principle, 
could be as late as shortly before BBN. 

In our scenario, however, it is the DM itself which gives rise to this modified expansion
history, whose ``kination'' phase is guaranteed due to its $U(1)$-symmetry ! 
Its dynamical evolution will ensure that the ``kination'' (stiff) phase will eventually go over into a radiation-like phase which can be 
brought into accordance with BBN and other observables, as shown in the previous sections. 
Also, the adopted Higgs boson masses in \cite{JP} were small and have been excluded by the 2012 measurement of the Higgs mass. 
A higher Higgs mass makes a strong enough 1st-order PT even more out of reach. 

We have embarked on the study of the implication of our modified expansion history due to SFDM in light of
the recently measured value of the Higgs mass, and detailed results will be published elsewhere.   
We adopt the model studied in \cite{Wells1,Wells2}, 
in which the (normally up to quartic) Higgs potential is augmented by a term of 6th power which is suppressed by a
cutoff scale $\Lambda \gtrsim 500$ GeV (the notation $\Lambda$ shall not be confused with the cosmological constant of previous 
sections).
This model is attractive, since it still accomodates the current bounds on Higgs properties, but leaves room for high-energy extensions 
to the standard model, which are encoded in the scale $\Lambda$. 
It was shown in \cite{Wells1, Wells2} that this model can facilitate a 1st-order PT in a standard
radiation-dominated Universe, depending on the value of $\Lambda$ and the (then unknown) Higgs mass. 
Now, we want to explore when
a sufficiently strong 1st-order EWPT could be achieved, given the known Higgs mass, depending on $\Lambda$ and the allowed SFDM models for the dark matter
with their modified expansion rates. 
In Figure \ref{fig3}, we show the critical and nucleation temperatures ($T_c, T_n$) on the left side, as well as the ``measure'' used to describe the 
effectiveness of a 1st-order PT, - 
the Higgs vev $v$ over $T_c$ or $T_n$- on the right side, each as a function of $\Lambda$ for different cosmologies. 

We stress two points: first, we can see that a modification of the cosmology (i.e. the change of the expansion rate due to the dominance of SFDM
over radiation) hardly affects the Higgs-particle model specifics of critical temperatures and $v/T$ as a function of $\Lambda$.
Second, the sphaleron freeze-out condition for $v/T$ does indeed become weaker (down from a value of $\geq 1.0$) for increasing expansion rates, at a given $\Lambda$.
Our EWSB era of interest lies between around a temperature of $70-150$ GeV (according to the left-hand plot of Figure 
\ref{fig3}). We highlight this epoch, as well as the BBN epoch, with vertical lines in the right-hand plot of 
Figure \ref{fig2}, which shows the Hubble expansion rate
as a function of scale factor during these epochs, for our fiducial SFDM model and $T_{\rm{re}} = 300$ GeV. We can see 
that the expansion rate during EWSB
is about 3-5 orders of magnitudes higher
with respect to $\Lambda$CDM, thanks to the stiff phase of $\Lambda$SFDM. The sphaleron bound gets accordingly weaker, $v/T \geq 0.73$ for $T_n = 70$ GeV or 
$v/T \geq 0.55$ for $T_n = 150$ GeV, respectively. 

Work is in progress to study the SFDM parameter space in more detail and
respective implications for $\Omega_{\rm{GW}}$, as well as EWSB. }

\Acknowledgments
{\small TRD acknowledges support by the U.S. Department of Energy under grants DE-FG02-95ER40899 and DE-SC0007859.
This work was supported in part by U.S. NSF grant AST-1009799, NASA grant NNX11AE09G, NASA/JPL grant RSA Nos.1492788
and 1515294, and supercomputer resources from NSF XSEDE grant TG-AST090005 and the Texas Advanced Computing Center (TACC)
at the University of Texas at Austin. ML was supported by the Polish National Science Centre under research grant
2014/13/N/ST2/02712 and doctoral scholarship 2015/16/T/ST2/00527. JW is supported in part by the Department of Energy under
grant DE-SC0007859.}

\end{document}